\RequirePackage{fix-cm}
\documentclass[twocolumn,epjc3]{svjour3}
\smartqed  
\RequirePackage{graphicx}
\journalname{Eur. Phys. J. C}
\usepackage{latexsym}
\usepackage{graphics}
\usepackage{bm}
\usepackage{mathrsfs}
\usepackage{dcolumn}
\usepackage[usenames,dvipsnames]{color}
\usepackage{eufrak}
\usepackage{lscape}
\usepackage{amsmath}
\begin{document}
\title{Localized direct CP violation in $B^\pm\rightarrow \rho^0 (\omega)\pi^\pm\rightarrow \pi^+ \pi^-\pi^\pm$}
\author{Chao Wang\thanksref{e1,addr1}
        \and
        Zhen-Hua Zhang\thanksref{e2,addr2}
        \and
        Zhen-Yang Wang\thanksref{e3,addr1}
        \and
        Xin-Heng Guo\thanksref{e4,addr1}
}
\thankstext{e1}{e-mail:chaowang11@mail.bnu.edu.cn}
\thankstext{e2}{e-mail:zhangzh@usc.edu.cn}
\thankstext{e3}{e-mail:rongtian@mail.bnu.edu.cn}
\thankstext{e4}{e-mail:Corresponding author: xhguo@bnu.edu.cn}


\institute{College of Nuclear Science and Technology, Beijing Normal University, Beijing, 100875, China \label{addr1}\and
School of Nuclear Science and Technology, University of South China, Hengyang, 421001, Hunan, China
\label{addr2}
}

\date{Received: date / Revised version: date}
%

\maketitle
\begin{abstract}
We study the localized direct CP violation in the hadronic decays $B^\pm\rightarrow \rho^0 (\omega)\pi^\pm\rightarrow\pi^+ \pi^-\pi^\pm$, including the effect caused by an interesting mechanism involving the charge symmetry violating mixing between $\rho^0$ and $\omega$. We calculate the localized integrated direct CP violation when the low invariant mass of $\pi^+\pi^-$ [$m(\pi^+\pi^-)_{low}$] is near $\rho^0(770)$. For five models of form factors investigated, we find that the localized integrated direct CP violation varies from -0.0752 to -0.0290 in the ranges of parameters in our model when $0.750<m(\pi^+\pi^-)_{low}<0.800$\,GeV. This result, especially the sign, agrees with the experimental data and is independent of form factor models. The new experimental data shows that the signs of the localized integrated CP asymmetries in the regions $0.470<m(\pi^+\pi^-)_{low}<0.770$\,GeV and $0.770<m(\pi^+\pi^-)_{low}<0.920$\,GeV are positive and negative, respectively. We find that $\rho$-$\omega$ mixing makes the localized integrated CP asymmetry move toward the negative direction, and therefore contributes to the sign change in those two regions. This behavior is also independent of form factor models. We also calculate the localized integrated direct CP violating asymmetries in the regions $0.470<m(\pi^+\pi^-)_{low}<0.770$\,GeV, $0.770<m(\pi^+\pi^-)_{low}<0.920$\,GeV and the whole region $0.470<m(\pi^+\pi^-)_{low}<0.920$\,GeV and find that they agree with the experimental data in some models of form factors.
\end{abstract}

\section{Introduction}
\label{intro}

CP violation is one of the most fundamental and important properties of weak interactions. Even though it has been known since 1964 \cite{J.C.}, we still do not know the source of CP violation completely. In the Standard Model, CP violation originates from the weak phase in the Cabibbo-Kobayashi-Maskawa (CKM) matrix \cite{N.C.,M.K.}. Besides the weak phase, a large strong phase is also needed for direct CP violation to occur in decay processes. Usually, this large phase is provided by QCD loop corrections and some phenomenological mechanisms. In the past few years, numerous theoretical studies have been conducted on CP violation. However, we need a lot of data to test these approaches because there are many theoretical uncertainties such as CKM matrix elements, hadronic matrix elements, and nonfactorizable effects. These uncertainties would be reduced by the increase of experimental data in the future and the improvement of theoretical methods.

Recently, the LHCb Collaboration focused on three-body final states in the decays of $B$ and $D$ mesons and a novel strategy to probe CP asymmetry in their Dalitz plots \cite{J.M.,R.A.1,R.A.2}. The local asymmetries in specific regions of the phase space of charmless three-body decays of bottom mesons, such as $B^\pm\to \pi^\pm \pi^+\pi^-$ and $B^\pm\to K^\pm \pi^+\pi^-$, were measured. It was shown that the local asymmetry distributions in the Dalitz plots reveal rich structures and are not uniform \cite{J.M.,R.A.1,R.A.2}. These intriguing discoveries offer opportunities to search for different sources of CP violation, through the study of the signatures of these sources in certain phase spaces of the Dalitz plots. In fact, several theoretical studies have been made to explain these distributions, such as the interference between intermediate states \cite{S.K.2,Z.H.,Z.H.1} and final-state rescatterings \cite{R.A.2,B.B.,I.B.,D.A.}. One can confirm that these complex structures originate from more than one source \cite{R.A.2}.

Charge symmetry is broken at the most fundamental level in strong interaction physics through the small mass difference between up and down quarks in the QCD Lagrangian. As a consequence the physical $\rho^0$ and $\omega$ mesons are not eigenstates of isospin but, for example, the physical $\rho^0$ contains a small admixture of an $I = 0$ state \cite{H.B.3}. In previous works, this phenomenon, known as $\rho$-$\omega$ mixing, was considered to obtain a large strong phase in $B$ decays \cite{R.E.,S.G.,X.H.1} and it was found that such a mixing can lead to a peak of CP violation when the invariant mass of $\pi^+\pi^-$ is near $\omega$. The differential CP asymmetry was studied in the decays $B^\pm\to\rho^0(\omega)\pi^\pm \to \pi^+\pi^-\pi^\pm$ before \cite{X.H.2,O.L.,G.L.}. $\rho$-$\omega$ mixing provides additional complex terms to the amplitudes \cite{C.W.} and the strong phase passes through $90\,^{\circ}$ at the $\omega$ resonance \cite{S.G.,X.H.1,X.H.2,O.L.,G.L.,G.L.2}. In our previous work, we found that it is more useful to investigate the localized integrated CP asymmetry and studied the localized integrated CP asymmetry in hadronic $\tau$ decays with this mechanism \cite{C.W.}. The newest LHCb experiments showed that the resonances are $\rho^0$(770) when $m^2(\pi^+\pi^-)_{low}<1\mathrm{GeV}^2/\mathrm{c}^4$ for $B^\pm\to\pi^+\pi^-\pi^\pm$ decays [$m(\pi^+\pi^-)_{low}$ is the low invariance mass of $\pi^+\pi^-$] \cite{R.A.2}. It was claimed that the CP asymmetry in these decays changes sign around the $\rho^0(770)$ peak of $m^2(\pi^+\pi^-)_{low}$ \cite{R.A.2}, which contains the significant region of $\rho$-$\omega$ mixing. In the present paper, we aim at studying the localized integrated CP asymmetry in $B^\pm\to\pi^+\pi^-\pi^\pm$ decays involving $\rho$-$\omega$ mixing and comparing it with the results obtained from the LHCb Collaboration.

In this paper, we will investigate the localized integrated CP violation with five phenomenological models of weak form factors for $B^\pm\rightarrow \rho^0(\omega)\pi^\pm\rightarrow \pi^-\pi^+\pi^\pm$ with and without $\rho$-$\omega$ mixing. We will determine the allowed range of $N_c$ which is the effective color number describing nonfactorizable contributions. The model dependence of our results will be discussed in detail. We will see that for five models of form factors, our result for the localized integrated CP asymmetry varies from -0.0752 (-0.0626) to -0.0403 (-0.0290) in the ranges of $N_c$, $2.07(2.09)<N_c<4.54(4.65)$ corresponding to $q^2/m_b^2=0.3(0.5)$ ($q$ is the typical momentum transfer of the gluon or photon in the penguin diagrams) when $0.750<m(\pi^+\pi^-)_{low}<0.800$\,GeV. From Fig.~4 of Ref.~\cite{R.A.2}, we can see the localized integrated CP asymmetries have different signs in $\cos\theta>0$ and $\cos\theta<0$ regions when $0.750<m(\pi^+\pi^-)_{low}<0.800\, \text{GeV}$ ($\theta$ is the angle between the momenta of the unpaired $\pi^{+(-)}$ and the $\rho^0$ decay product with the same-sign charge). If one adds the events in these two experimental regions together, the total localized integrated CP asymmetry will be $-0.0294\pm0.0285$. Our results agree with this experimental data. The experimental values of the localized integrated CP asymmetries in the regions $0.470<m(\pi^+\pi^-)_{low}<0.770\, \text{GeV}$ and $0.770<m(\pi^+\pi^-)_{low}<0.920\, \text{GeV}$ are $0.0508\pm0.0171$ and $-0.0256\pm0.0202$, respectively, with opposite signs \cite{R.A.2}. We will find that $\rho$-$\omega$ mixing can make the localized integrated CP asymmetry move toward the negative direction, and therefore will contribute to the sign change from the region $0.470<m(\pi^+\pi^-)_{low}<0.770\, \text{GeV}$ to $0.770<m(\pi^+\pi^-)_{low}<0.920\, \text{GeV}$. This behavior is independent of form factor models. Furthermore, we will find that our results in these two regions and the whole region of them are consistent with the experimental data for several models of form factors.

The remainder of this paper is organized as follows. In Sect. 2, we present the form of the effective Hamiltonian and the values of Wilson coefficients and give the formalism for the CP violating asymmetry in $B^\pm\rightarrow \rho^0(\omega)\pi^\pm\rightarrow \pi^+\pi^-\pi^\pm$. Then we show numerical results with several models of form factors in this section. In Sect.~3, we calculate branching ratios for $B^+\to\rho^0\pi^+$ and $B^0\to \rho^+\pi^-$ and present numerical results for the range of $N_c$ allowed by the experimental data. In the last section, we give some discussions and summarize our results.

\section{CP violation in $B^\pm\rightarrow \rho^0 (\omega)\pi^\pm\rightarrow \pi^+ \pi^-\pi^\pm$}
The amplitude of a decay process described by some amplitudes may have CP-even and -odd relative phases. Within the Standard Model, the CP-odd relative phase is always a weak phase difference which is directly determined by the CKM matrix. On the other hand, CP-even phases, which are called strong phases, usually originates from nonperturbative effects of strong interactions and are hard to handle. We consider a $B$ meson weak decay process, $B\to M_1 M_2 M_3$, where $M_i(i=1,2,3)$ is a light pseudoscalar meson. For a weak decay process of a heavy meson, a typical form of the decay amplitude $A$ and its CP conjugate one $\bar A$ are
\begin{eqnarray}
A&=&g_1r_1\mathrm{e}^{\mathrm{i}\phi_1}+g_2r_2\mathrm{e}^{\mathrm{i}\phi_2},\label{M}\\
\bar A&=&g_1^*r_1\mathrm{e}^{\mathrm{i}\phi_1}+g_2^*r_2\mathrm{e}^{\mathrm{i}\phi_2},\label{Mbar}
\end{eqnarray}
where $g_1$ and $g_2$ represent CP-odd complex terms which involve CKM matrix elements, $r_1e^{\mathrm{i}\phi_1}$ and $r_2e^{\mathrm{i}\phi_2}$ terms are even under the CP transformation. Then one has
\begin{eqnarray}
\label{M-M}
&&|A|^2-|\bar A|^2=4r_1r_2\,\mathfrak{I}\mathfrak{m}(g_1^*g_2)\,\sin(\phi_1-\phi_2)\nonumber\\
&&\quad=4r_1r_2|g_1||g_2|\,\sin [\,\mathrm{Arg}(g_2/g_1)]\,\sin(\phi_1-\phi_2),
\end{eqnarray}
from which, we can see explicitly that both the CP-odd phase difference $\mathrm{Arg}(g_2/g_1)$ and the CP-even phase difference $\phi_1-\phi_2$ are needed to produce CP violation.
\subsection{The effective Hamiltonian}

In order to calculate the direct CP violating asymmetry in hadronic decays, one can use the following effective weak Hamiltonian, based on the Operator Product Expansion \cite{G.B.,M.B.1,M.B.2,M.B.3,M.B.4}:
\begin{equation}
\mathcal{H}=\frac{G_F}{\sqrt 2}\big[V_{ub}V^*_{ud}(c_1O^u_1+c_2O^u_2)-V_{tb}V^*_{td}\sum^{10}_{i=3}c_iO_i\big]+H.c.,
\end{equation}
where $V_{ub}$, $V_{ud}$, $V_{tb}$ and $V_{td}$ are CKM matrix elements, and $c_i(i=1,2,..,10)$ are the Wilson coefficients, which are calculable in the renormalization group improved perturbation theory and are scale dependent. In the present case, we work with the renormalization scheme independent Wilson coefficients and use the values of the Wilson coefficients at the renormalization scale $\mu\approx m_b$. The operators $O_i$ have the following forms:
\begin{eqnarray}{\label O}
O^u_1&=&\overline{d}_\alpha\gamma_\mu(1-\gamma_5)u_\beta\overline{u}_\beta\gamma^\mu(1-\gamma_5)b_\alpha,\nonumber\\
O^u_2&=&\overline{d}\gamma_\mu(1-\gamma_5)u\overline{u}\gamma^\mu(1-\gamma_5)b,\nonumber\\
O_3&=&\overline{d}\gamma_\mu(1-\gamma_5)b\sum_{q^\prime}\overline{q^\prime}\gamma^\mu(1-\gamma_5)q^\prime,\nonumber\\
O_4&=&\overline{d}_\alpha\gamma_\mu(1-\gamma_5)b_\beta\sum_{q^\prime}\overline{q^\prime}_\beta\gamma^\mu(1-\gamma_5)q^\prime_\alpha,\nonumber\\
O_5&=&\overline{d}\gamma_\mu(1-\gamma_5)b\sum_{q^\prime}\overline{q^\prime}\gamma^\mu(1+\gamma_5)q^\prime,\nonumber\\
O_6&=&\overline{d}_\alpha\gamma_\mu(1-\gamma_5)b_\beta\sum_{q^\prime}\overline{q^\prime}_\beta\gamma^\mu(1+\gamma_5)q^\prime_\alpha,\nonumber\\
O_7&=&\frac{3}{2}\overline{d}\gamma_\mu(1-\gamma_5)b\sum_{q^\prime}e_{q^\prime}\overline{q^\prime}\gamma^\mu(1+\gamma_5)q^\prime,\nonumber\\
O_8&=&\frac{3}{2}\overline{d}_\alpha\gamma_\mu(1-\gamma_5)b_\beta\sum_{q^\prime}e_{q^\prime}\overline{q^\prime}_\beta\gamma^\mu(1+\gamma_5)q^\prime_\alpha,\nonumber\\
O_9&=&\frac{3}{2}\overline{d}\gamma_\mu(1-\gamma_5)b\sum_{q^\prime}e_{q^\prime}\overline{q^\prime}\gamma^\mu(1-\gamma_5)q^\prime,\nonumber\\
O_{10}&=&\frac{3}{2}\overline{d}_\alpha\gamma_\mu(1-\gamma_5)b_\beta\sum_{q^\prime}e_{q^\prime}\overline{q^\prime}_\beta\gamma^\mu(1-\gamma_5)q^\prime_\alpha,
\end{eqnarray}
where $\alpha$ and $\beta$ are color indices, and $q^\prime=u, d$ or $s$ quarks. In Eq.~(\ref{O}), $O_1^u$ and $O_2^u$ are the tree level operators, $O_3-O_6$ are QCD penguin operators, and $O_7-O_{10}$ arise from electroweak penguin diagrams.

The Wilson coefficients, $c_i$, which are known to the next-to-leading logarithmic order, take the following values \cite{N.G.,R.F.}:

\begin{eqnarray}
c_1&=&-0.3125,\qquad c_2=1.1502,\nonumber\\
c_3&=&0.0174,\qquad c_4=-0.0373,\nonumber\\
c_5&=&0.0104,\qquad c_6=-0.0459,\nonumber\\
c_7&=&-1.050\times10^{-5},\qquad c_8=3.839\times10^{-4},\nonumber\\
c_9&=&-0.0101,\qquad  c_{10}=1.959\times10^{-3},
\end{eqnarray}
at the scale $\mu= m_b=5\,\text{GeV}$.

To be consistent, the matrix elements of the operators $O_i$ should also be renormalized to the one-loop order. This results in the effective Wilson coefficients, $c^\prime_i$, which satisfy the constraint

\begin{equation}{\label{hr}}
c_i(m_b)\langle O_i (m_b) \rangle=c^\prime_i\langle O_i\rangle^{tree},
\end{equation}
where $\langle O_i \rangle ^{tree}$ is the matrix element at the tree level, which will be evaluated in the factorization approach. From Eq.~(\ref{hr}), the relations between $c^\prime_i$ and $c_i$ are \cite{N.G.,R.F.}
\begin{eqnarray}
c_1^\prime&=&c_1,\qquad c_2^\prime=c_2,\nonumber\\
c_3^\prime&=&c_3-P_s/3,\qquad c_4^\prime=c_4+P_s,\nonumber\\
c_5^\prime&=&c_5-P_s/3,\qquad c_6^\prime=c_6+P_s,\nonumber\\
c_7^\prime&=&c_7+P_e,\qquad c_8^\prime=c_8,\nonumber\\
c_9^\prime&=&c_9+P_e,\qquad c_{10}^\prime=c_{10},
\end{eqnarray}
where
\begin{eqnarray}
P_s&=&(\alpha_s/8 \pi)c_2[10/9+G(m_c,\mu,q^2)],\nonumber\\
P_e&=&(\alpha_{em}/9 \pi)(3c_1+c_2)[10/9+G(m_c,\mu,q^2)],\nonumber
\end{eqnarray}
with
\begin{eqnarray}
G(m_c,\mu,q^2)=4\int^1_0\text{d}xx(x-1)\text{ln}\frac{m^2_c-x(1-x)q^2}{\mu^2}.\nonumber
\end{eqnarray}
$G(m_c,\mu,q^2)$ has the following explicit expression \cite{G.K.}:
\begin{eqnarray}
\mathfrak{R}\mathfrak{e}G&=&\frac{2}{3}\Bigg[\text{ln}\frac{m_c^2}{\mu^2}-\frac{5}{3}-4\frac{m_c^2}{q^2}+\big(1+2\frac{m_c^2}{q^2}\big)\nonumber\\
&&\qquad\times\sqrt{1-4\frac{m_c^2}{q^2}}\text{ln}\frac{1+\sqrt{1-4\frac{m_c^2}{q^2}}}{1-\sqrt{1-4\frac{m_c^2}{q^2}}}\Bigg],\nonumber\\
\mathfrak{I}\mathfrak{m}G&=&-\frac{2}{3}\Big(1+2\frac{m_c^2}{q^2}\Big)\sqrt{1-4\frac{m_c^2}{q^2}}.
\end{eqnarray}
Based on simple arguments at the quark level, the value of $q^2$ is chosen to be in the range $0.3<q^2/m^2_b<0.5$ \cite{R.E.}. From Eqs.~(8) and (9) we can obtain numerical values for $c^\prime_i$.

When $q^2/m_b^2=0.3$,
\begin{eqnarray}
c^\prime_1&=&-0.3125,\qquad c^\prime_2=1.1502,\nonumber\\
c^\prime_3&=&2.433\times10^{-2}+1.543\times10^{-3}\text{i},\nonumber\\
c^\prime_4&=&-5.808\times10^{-2}-4.628\times10^{-3}\text{i},\nonumber\\
c^\prime_5&=&1.733\times10^{-2}+1.543\times^{-3}\text{i},\nonumber\\
c^\prime_6&=&-6.668\times10^{-2}-4.628\times^{-3}\text{i},\nonumber\\
c^\prime_7&=&-1.435\times10^{-4}-2.963\times^{-5}\text{i},\nonumber\\
c^\prime_8&=&3.839\times10^{-4},\nonumber\\
c^\prime_9&=&-1.023\times10^{-2}-2.963\times^{-5}\text{i},\nonumber\\
c^\prime_{10}&=&1.959\times10^{-3},
\end{eqnarray}
and when $q^2/m_b^2=0.5$, one has
\begin{eqnarray}
c^\prime_1&=&-0.3125,\qquad c^\prime_2=1.1502,\nonumber\\
c^\prime_3&=&2.120\times10^{-2}+2.174\times10^{-3}\text{i},\nonumber\\
c^\prime_4&=&-4.869\times10^{-2}-1.552\times10^{-3}\text{i},\nonumber\\
c^\prime_5&=&1.420\times10^{-2}+5.174\times^{-3}\text{i},\nonumber\\
c^\prime_6&=&-5.729\times10^{-2}-1.552\times^{-2}\text{i},\nonumber\\
c^\prime_7&=&-8.340\times10^{-5}-9.938\times^{-5}\text{i},\nonumber\\
c^\prime_8&=&3.839\times10^{-4},\nonumber\\
c^\prime_9&=&-1.017\times10^{-2}-9.938\times^{-5}\text{i},\nonumber\\
c^\prime_{10}&=&1.959\times10^{-3},
\end{eqnarray}
where we have taken $\alpha_s(m_Z)=0.112$, $\alpha_{em}(m_b)=1/132.2$, $m_b=5$\,GeV, and $m_c=1.35$\,GeV.

\subsection{Formalism}

For the $B^-\rightarrow\rho^0\pi^-$ process, the amplitude can be written as $M_{B^-\rightarrow\rho^0\pi^-}^\lambda=\alpha p_B\cdot\epsilon^*(\lambda)$, where $\epsilon$ is the polarization vector of $\rho^0$ and $\lambda$ is its polarization, $p_B$ is the momenta of $B^-$ meson, and $\alpha$ is independent of $\lambda$. The amplitude for $\rho^0\rightarrow\pi^+\pi^-$ is $M_{\rho^0\rightarrow\pi^+\pi^-}^\lambda=g_\rho \epsilon(\lambda)(p_1-p_2)$, where $p_1$ and $p_2$ are the momenta of $\pi^+$ and $\pi^-$ produced by $\rho^0$, respectively, and $g_\rho$ is the effective coupling constant for $\rho^0\rightarrow\pi^+\pi^-$. Then, for the sequential decay $B^-\rightarrow\rho^0\pi^-\rightarrow\pi^+\pi^-\pi^-$, the amplitude is \cite{Z.H.,Z.H.1}:
\begin{eqnarray}
 A&=&\alpha p_B^\mu \frac{\sum_\lambda\epsilon^*_\mu(\lambda)\epsilon_\nu(\lambda)}{s_\rho}g_\rho(p_1-p_2)^\nu \nonumber\\
 &=&\frac{g_\rho\alpha}{s_\rho}\cdot p_B^\mu\Big[ g_{\mu\nu}-\frac{(p_1+p_2)_\mu(p_1+p_2)_\nu}{s} \Big](p_1-p_2)^\nu \nonumber\\
 &=&\frac{g_\rho}{s_\rho}\cdot \frac{M_{B^+\rightarrow\rho^0\pi^+}^\lambda}{p_B\cdot\epsilon^*}\cdot (\Sigma-s^\prime)\nonumber\\
 &=&(\Sigma-s^\prime)\cdot \mathcal{M},
\end{eqnarray}
where $\sqrt {s^\prime}$ is the high invariance mass of the $\pi^+\pi^-$ pair, $\Sigma=\frac{1}{2}(s^\prime_{max}+s^\prime_{min})$ with $s^\prime_{max}$ and $s^\prime_{min}$ being the maximum and minimum values of $s^\prime$ for a fixed $s$, respectively, and $\sqrt s$ is the low invariant mass of the $\pi^+\pi^-$ pair [$m(\pi^+\pi^-)_{low}$] and $s_V$ is from the inverse propagator of the vector meson $V$, $s_V=s-m_V^2+\mathrm{i}m_V\Gamma_V $. The factor $\Sigma-s^\prime$ is equal to $-2|\vec{p_2}||\vec{p_3}|\cos{\theta}$ in Ref.~\cite{I.B.0} , which accounts for angular momentum conservation for the spin-1 resonance. $\mathcal{M}$ will be calculated in the following. According to the effective Hamiltonian, $A$ can be divided into two parts \cite{X.H.2}:
\begin{eqnarray}
A&=&\langle\pi^+\pi^-\pi^-|H^T|B^-\rangle+\langle\pi^+\pi^-\pi^-|H^P|B^-\rangle,
\end{eqnarray}
where $H^T$ and $H^P$ are the Hamiltonians for the tree and penguin operators, respectively.

In order to obtain a large signal for direct CP violation, we need to appeal to some phenomenological mechanisms. $\rho$-$\omega$ mixing has the dual advantages that the strong phase difference is large (passes through $90\,^{\circ}$ at the $\omega$ resonance) and well known \cite{S.G.,X.H.1}. With this mechanism, to the first order in isospin violation, the amplitude for $B^-\to \rho^0(\omega)\pi^- \to \pi^+\pi^-\pi^-$ takes the following form at a value of $\sqrt s$ close to the $\omega$ resonance mass \cite{X.H.2}:
\begin{eqnarray}
\langle\pi^+\pi^-\pi^-|H^T|B^-\rangle&=&(\Sigma-s^\prime)\nonumber\\
&&\times\left(\frac{g_\rho}{s_\rho s_\omega}\tilde{\Pi}_{\rho \omega}t_\omega+\frac{g_\rho}{s_\rho}t_{\rho}\right),\\
\langle\pi^+\pi^-\pi^-|H^P|B^-\rangle&=&(\Sigma-s^\prime)\nonumber\\
&&\times\left(\frac{g_\rho}{s_\rho s_\omega}\tilde{\Pi}_{\rho \omega}p_\omega+\frac{g_\rho}{s_\rho}p_{\rho}\right),
\end{eqnarray}
where $t_V$ ($V=\rho^0$ or $\omega$) is the tree amplitude and $p_V$ is the penguin amplitude for producing an intermediate vector meson $V$, $\tilde{\Pi}_{\rho \omega}$ is the effective $\rho$-$\omega$ mixing amplitude. From Eqs.~(14) and (15), we note that $\rho$-$\omega$ mixing provides an additional complex term for the tree and penguin amplitudes (the first term in each equation), respectively. These complex terms will enlarge the CP-even phase and lead to a peak of CP asymmetry as mentioned before. We will show the difference between the CP asymmetries with and without $\rho$-$\omega$ mixing later. Here, we assume that the $B^\pm\to \pi^+\pi^-\pi^\pm$ process is dominated by the resonance $\rho^0$ in certain region of its Dalitz plot.

We stress that the direct coupling $\omega\to\pi^+\pi^-$ is effectively absorbed into $\tilde{\Pi}_{\rho \omega}$ \cite{H.B.}, leading to the explicit $s$ dependence of $\tilde{\Pi}_{\rho \omega}$. Making the expansion $\tilde{\Pi}_{\rho \omega}(s)=\tilde{\Pi}_{\rho \omega}(m_\omega^2)+(s-m_\omega^2)\tilde{\Pi}_{\rho \omega}^\prime(m_\omega^2)$, the $\rho$-$\omega$ mixing parameters were determined in the fit of Gardner and O'Connell \cite{S.G.2}:
\begin{eqnarray}
\label{mix}
\mathfrak{R} \mathfrak{e}\tilde{\Pi}_{\rho \omega}(m^2_{\omega})&=&{}-3500\pm 300\ \mathrm{MeV^2},\nonumber\\
\mathfrak{I} \mathfrak{m}\tilde{\Pi}_{\rho \omega}(m^2_{\omega})&=&{}-300\pm 300\ \mathrm{MeV^2},\nonumber\\
\tilde{\Pi}_{\rho \omega}^\prime(m^2_{\omega})&=&0.03\pm\ 0.04.
\end{eqnarray}
In practice, the effect of the derivative term is negligible.

In this work, we only consider $\rho^0$ and $\omega$ resonances. Then, for a fixed $s$, the differential CP asymmetry parameter can be defined as
\begin{eqnarray}
A_{CP}=\frac{|\mathcal{M}|^2-|\bar{\mathcal{M}}|^2}{|\mathcal{M}|^2+|\bar{\mathcal{M}}|^2}.
\end{eqnarray}
By integrating the denominator and numerator of $A_{CP}$, respectively, in the region $\Omega$ ($s_1<s<s_2$, $s^\prime_1<s^\prime<s^\prime_2$), we obtain the localized integrated CP asymmetry, which can be measured by experiments and takes the following form:
\begin{eqnarray}
A_{CP}^\Omega = \frac{\int_{s_1}^{s_2} \mathrm{d} s   \int_{s^\prime_1}^{s^\prime_2} \mathrm{d}s^\prime (\Sigma-s^\prime)^2 (|\mathcal{M}|^2-\bar{|\mathcal{M}|}^2)}{\int_{s_1}^{s_2} \mathrm{d} s   \int_{s^\prime_1}^{s^\prime_2} \mathrm{d}s^\prime(\Sigma-s^\prime)^2 (|\mathcal{M}|^2+\bar{|\mathcal{M}|}^2)}.\label{acpin}
\end{eqnarray}
According to kinematics of the three body decay, $\Sigma[=\frac{1}{2}(s^\prime_{max}+s^\prime_{min})]$ is related to $s$. In our calculations, $s$ varies in a small region, and therefore $\Sigma$ can be treated as a constant approximately \cite{R.A.2}. Then, the terms $\int_{s^\prime_1}^{s^\prime_2} \mathrm{d} s^\prime(\Sigma-s^\prime)^2$ are cancelled, and $A_{CP}^\Omega$ becomes independent of the high invariance mass of $\pi^+\pi^-$. In practice, to be more precise, we take into account the $s$-dependence of $s^\prime_{max}$ and $s^\prime_{min}$ in our calculations. We choose $s^\prime_{min}<s^\prime<s^\prime_{max}$ as the integration interval of the high invariance mass of $\pi^+\pi^-$ and regard $\int_{s^\prime_{min}}^{s^\prime_{max}} \mathrm{d} s^\prime(\Sigma-s^\prime)^2$ as a factor which is dependent on $s$.

\subsection{Calculational details}

With the Hamiltonian given in Eq.~(4), we are ready to evaluate the matrix elements for $B^-\to\rho^0(\omega)\pi^-$. According to the theory of QCD factorization, the naive factorization approach has been shown to be the leading order result in the framework of QCD factorization when the radiative QCD corrections of order $\mathcal O(\alpha_s(m_b))$ and the $\mathcal O(1/m_b)$ corrections in the heavy quark effective theory are neglected \cite{M.B.1,M.B.2,M.B.3,M.B.4}. Since the $b$ quark is very heavy and $B$ meson decays are very energetic, so the quark-antiquark pair in a meson in the final state moves very fast away from the weak interaction point. The hadronization of the quark-antiquark pair occurs far away from the remaining quarks. Then the meson can be factorized out and the interaction between the quark pair in the meson and the remaining quark is tiny \cite{J.D.,M.J.}. The deviation of the value of $N_c$ from the color number, 3, measures the nonfactorizable effects in the naive factorization scheme \cite{X.H.1,X.H.2}. In the factorization approximation, either $\rho^0(\omega)$ or $\pi^-$ is generated by one current which has appropriate quantum numbers in the Hamiltonian. For this decay process, two kinds of matrix element products are involved after factorization: $\langle\rho^0(\omega)|J^\mu|0\rangle\langle\pi^-|J_\mu|B^-\rangle$ and $\langle\pi^-|J^\mu|0\rangle \langle\rho^0(\omega)|J_\mu|B^-\rangle$. We will calculate them in some phenomenological quark models.


The matrix elements for $B\to P$ and $B\to V$ (where $P$ and $V$ denote pseudoscalar and vector mesons, respectively) can be decomposed as \cite{M.W.}
\begin{eqnarray}
\langle P| J_\mu|B\rangle &=& \Big( p_B+p_P-\frac{m_B^2-m_P^2}{k^2}\Big)_\mu F_1^{BP}(k^2)\nonumber\\
&&+\frac{m_B^2-m_P^2}{k^2}k_\mu F_0^{BP}(k^2),\\
\langle V|J_\mu|B\rangle&=&\frac{2}{m_B+m_{V}}\varepsilon_{\mu\nu\rho\sigma}\epsilon^{*\nu}p_B^\rho p^\sigma_{V}V^{BV}(k^2)\nonumber\\
&&+\text{i}\Big\{\epsilon^*_\mu(m_B+m_{V})A_1^{BV}(k^2) \nonumber \\
&&\qquad-\frac{\epsilon^* \cdot k}{m_B+m_{V}}(p_B+p_{V})_\mu A_2^{BV} (k^2)\nonumber\\
&&\qquad-\frac{\epsilon^*\cdot k}{k^2}2m_{V}\cdot k_\mu A_3^{BV}(k^2)\Big\}\nonumber \\
&&+\text{i}\frac{\epsilon^*\cdot k}{k^2}2m_{V}\cdot k_\mu A^{BV}_0(k^2),
\end{eqnarray}
where $J_\mu$ is the weak current [$J_\mu=\bar u \gamma_\mu (1-\gamma_5)b$ or $\bar d \gamma_\mu (1-\gamma_5)b$], $k=p_B-p_{P(V)}$, and $\epsilon_\mu$ is the polarization vector of $V$. The form factors included in our calculations satisfy $F_1^{BP}(0)=F_0^{BP}(0)$, $A_3^{BV}(0)=A_0^{BV}(0)$, and $A_3^{BV}(k^2)=[(m_B+m_{V})/2m_{V}]A_1^{BV}(k^2)-[(m_B-m_{V})/2m_{V}]A_2^{BV}(k^2)$. We define the notation $X$ for matrix elements. For example, $X^{(B^-\rho^0,\pi^-)}$ is defined as $\langle\pi^-|\bar d \gamma^\mu (1-\gamma_5) u|0\rangle\langle\rho^0|\bar u \gamma_\mu (1-\gamma_5)b|B^-\rangle$. These matrix elements can be parameterized as the products of decay constants and form factors. Therefore, the factorized terms $X^{(BM_1,M_2)}$ have the following expressions \cite{Y.H.}:
\begin{eqnarray}
X^{(BP,V)}&=&\langle V|\bar q_2 \gamma^\mu (1-\gamma_5)q_3|0\rangle\langle P|\bar q_1 \gamma_\mu (1-\gamma_5)b|B\rangle \nonumber\\
&=&2f_V m_V F_1^{BP}(m^2_V)(\epsilon^*\cdot p_B), \\
X^{(BV,P)}&=&\langle P|\bar q_2 \gamma^\mu (1-\gamma_5)q_3|0\rangle\langle V|\bar q_1 \gamma_\mu (1-\gamma_5)b|B\rangle \nonumber\\
&=&2f_P m_V A_0^{BV}(m^2_P)(\epsilon^*\cdot p_B),
\end{eqnarray}
where $f_V$ and $f_P$ are the decay constants of vector and pseudoscalar mesons, respectively.
Using the decomposition in Eqs.~(19)-(22), one has
\begin{eqnarray}
t_\rho&=&V_{ub}V^*_{ud}\Bigg(a_1\frac{X^{(B^-\rho^0,\pi^-)}}{\epsilon^*\cdot p_B}+a_2\frac{X^{(B^-\pi^-,\rho^0)}}{\epsilon^*\cdot p_B}\Bigg),\\
t_\omega&=&V_{ub}V^*_{ud}\Bigg(a_1\frac{X^{(B^-\omega,\pi^-)}}{\epsilon^*\cdot p_B}+a_2\frac{X^{(B^-\pi^-,\omega)}}{\epsilon^*\cdot p_B}\Bigg),
\end{eqnarray}
where all the $a_i$ are built up from the effective Wilson coefficients $c_i^\prime$'s, and take the form $a_i=c^\prime_i+c^\prime_{i+1}/N_c$ for odd $i$ and $a_i=c^\prime_i+c^\prime_{i-1}/N_c$ for even $i$. It is noted that in the factorization approach $N_c$ includes nonfactorizable contributions effectively, and the value of $N_c$ should be determined by experiments since we cannot evaluate nonfactorizable contributions. In the same way, we obtain the penguin operator contributions:
\begin{eqnarray}
p_\rho&=&-V_{tb}V^*_{td}\Bigg\{\Big[-a_4+\frac{3}{2}a_7+\frac{3}{2}a_9+\frac{1}{2}a_{10}\Big]\frac{X^{(B^-\pi^-,\rho^0)}}{\epsilon^*\cdot p_B}\nonumber\\
&&+\Big[a_4+a_{10}-2(a_6+a_8)\frac{m^2_\pi}{(m_d+m_u)(m_b+m_u)}\Big]\nonumber\\
&&\qquad\times\frac{X^{(B^-\rho^0,\pi^-)}}{\epsilon^*\cdot p_B} \Bigg\},\\
p_\omega&=&-V_{tb}V^*_{td}\Bigg\{\Big[2a_3+a_4+2a_5+\frac{1}{2}(a_7+a_9-a_{10})\Big]\nonumber\\
&&\qquad\qquad\times\frac{X^{(B^-\pi^-,\omega)}}{\epsilon^*\cdot p_B}\nonumber\\
&&+\Big[a_4+a_{10}-2(a_6+a_8)\frac{m^2_\pi}{(m_d+m_u)(m_b+m_u)}\Big]\nonumber\\
&&\qquad\qquad\times\frac{X^{(B^-\omega,\pi^-)}}{\epsilon^*\cdot p_B} \Bigg\}.
\end{eqnarray}
We adopt the same decay constants and form factors for the matrix elements producing $\rho^0$ and $\omega$ mesons. Then we have $X^{(B^-\rho^0,\pi^-)}=X^{(B^-\omega,\pi^-)}=2f_\pi m_\rho A_0(m_\pi^2)(\epsilon^* \cdot p_B)$ and $X^{(B^-\pi^-,\rho^0)}=X^{(B^-\pi^-,\omega)}=\sqrt 2 f_\rho m_\rho F_1(m_\rho^2)(\epsilon^* \cdot p_B)$, where \\$\langle\rho^0(\omega)|J^\mu|0\rangle=1/\sqrt2f_\rho m_\rho \epsilon^{*\mu}$ and $\langle\pi^-|J_\mu|0\rangle=\text{i}f_\pi p_\mu$.

\subsection{Numerical results}

\begin{table*}[htb]
\centering
\caption{The localized integrated asymmetries $A^\Omega_{CP}$ when $q^2/m_b^2=0.3(0.5)$. The first and second lines of each model corresponds to $A^\Omega_{CP}$ with and without $\rho$-$\omega$ mixing in the region $0.750<\sqrt s<0.800$\,GeV, respectively. The third and forth lines of each model correspond to the low-mass region ($0.470<\sqrt s<0.770$\,GeV) and the high-mass region ($0.770<\sqrt s<0.920$\,GeV) near the resonance mass, respectively. The fifth and sixth lines of each model corresponds to $A^\Omega_{CP}$ with and without $\rho$-$\omega$ mixing in the region $0.470<\sqrt s<0.920$\,GeV, respectively. The second and third columns correspond to $N_c=2.07(2.09)$. The fourth and fifth columns correspond to $N_c=4.54(4.65)$. The second and fourth columns correspond to lower limiting values of the CKM matrix elements. The third and fifth columns correspond to upper limiting values of the CKM matrix elements.}
\begin{tabular*}{16cm}{@{\extracolsep{\fill}}ccccc}
\hline
\hline
$N_c$          &\multicolumn{2}{c}{2.07(2.09)}&\multicolumn{2}{c}{4.54(4.65)}\\
$\rho,\eta$    &min  &  max                   &min  &  max              \\
\hline
Model 1\\
$0.750<\sqrt s<0.800$\,GeV&-0.0647(-0.0520)&-0.0736(-0.0591)&-0.0455(-0.0300)&-0.0517(-0.0341)\\
&0.0054(0.0204)&0.0062(0.0232)&0.0079(0.0281)&0.0090(0.0319)\\
$0.470<\sqrt s<0.770$\,GeV&-0.0024(0.0061)&-0.0027(0.0070)&0.0017(0.0278)&0.0019(0.0290)\\
$0.770<\sqrt s<0.920$\,GeV&-0.0352(-0.0128)&-0.0400(-0.0145)&-0.0239(0.0014)&-0.0272(0.0016)\\
$0.470<\sqrt s<0.920$\,GeV&-0.0167(-0.0021)&-0.0189(-0.0023)&-0.0093(0.0096)&-0.0106(0.0109)\\
&0.0054(0.0204)&0.0062(0.0232)&0.0079(0.0281)&0.0090(0.0319)\\

Model 2\\
$0.750<\sqrt s<0.800$\,GeV&-0.0661(-0.0516)&-0.0752(-0.0587)&-0.0469(-0.0290)&-0.0533(-0.0330)\\
&0.0062(0.0234)&0.0071(0.0267)&0.0092(0.0327)&0.0105(0.0371)\\
$0.470<\sqrt s<0.770$\,GeV&-0.0019(0.0086)&-0.0021(0.0098)&0.0026(0.0306)&0.0030(0.0353)\\
$0.770<\sqrt s<0.920$\,GeV&-0.0359(-0.0112)&-0.0408(-0.0127)&-0.0247(0.0038)&-0.0280(0.0043)\\
$0.470<\sqrt s<0.920$\,GeV&-0.0166(0.00004)&-0.0189(0.00004)&-0.0091(0.0129)&-0.0103(0.0146)\\
&0.0062(0.0234)&0.0071(0.0267)&0.0092(0.0327)&0.0105(0.0371)\\

Model 3\\
$0.750<\sqrt s<0.800$\,GeV&-0.0647(-0.0520)&-0.0737(-0.0592)&-0.0455(-0.0300)&-0.0515(-0.0341)\\
&0.0054(0.0204)&0.0062(0.0232)&0.0079(0.0281)&0.0090(0.0320)\\
$0.470<\sqrt s<0.770$\,GeV&-0.0024(0.0061)&-0.0028(0.0069)&0.0017(0.0278)&0.0019(0.0292)\\
$0.770<\sqrt s<0.920$\,GeV&-0.0350(-0.0125)&-0.0398(-0.0142)&-0.0238(0.0016)&-0.0270(0.0016)\\
$0.470<\sqrt s<0.920$\,GeV&-0.0167(-0.0021)&-0.0190(-0.0023)&-0.0094(0.0097)&-0.0106(0.0110)\\
&0.0054(0.0204)&0.0062(0.0232)&0.0079(0.0281)&0.0090(0.0320)\\

Model 4\\
$0.750<\sqrt s<0.800$\,GeV&-0.0661(-0.0517)&-0.0752(-0.0587)&-0.0469(-0.0290)&-0.0534(-0.0330)\\
&0.0062(0.0235)&0.0071(0.0267)&0.0092(0.0327)&0.0105(0.0372)\\
$0.470<\sqrt s<0.770$\,GeV&-0.0019(0.0086)&-0.0022(0.0098)&0.0026(0.0306)&0.0030(0.0353)\\
$0.770<\sqrt s<0.920$\,GeV&-0.0357(-0.0110)&-0.0405(-0.0125)&-0.0245(0.0040)&-0.0278(0.0046)\\
$0.470<\sqrt s<0.920$\,GeV&-0.0167(0.00005)&-0.0190(0.00005)&-0.0091(0.0129)&-0.0104(0.0146)\\
&0.0062(0.0235)&0.0071(0.0267)&0.0092(0.0327)&0.0105(0.0372)\\

Model 5\\
$0.750<\sqrt s<0.800$\,GeV&-0.0577(-0.0550)&-0.0658(-0.0626)&-0.0403(-0.0375)&-0.0459(-0.0427)\\
&0.0011(0.0041)&0.0012(0.0047)&0.0015(0.0055)&0.0017(0.0062)\\
$0.470<\sqrt s<0.770$\,GeV&-0.0055(-0.0083)&-0.0063(-0.0094)&-0.0034(-0.0046)&-0.0039(-0.0053)\\
$0.770<\sqrt s<0.920$\,GeV&-0.0322(-0.0216)&-0.0366(-0.0245)&-0.0222(-0.122)&-0.0252(-0.0139)\\
$0.470<\sqrt s<0.920$\,GeV&-0.0170(-0.0140)&-0.0194(-0.0160)&-0.0114(-0.0079)&-0.0130(-0.0090)\\
&0.0010(0.0039)&0.0012(0.0044)&0.0015(0.0052)&0.0017(0.0059)\\

\hline
\hline
\end{tabular*}
\end{table*}

\begin{figure*}
\scalebox{0.78}[0.78]{\includegraphics[10,266][489,559]{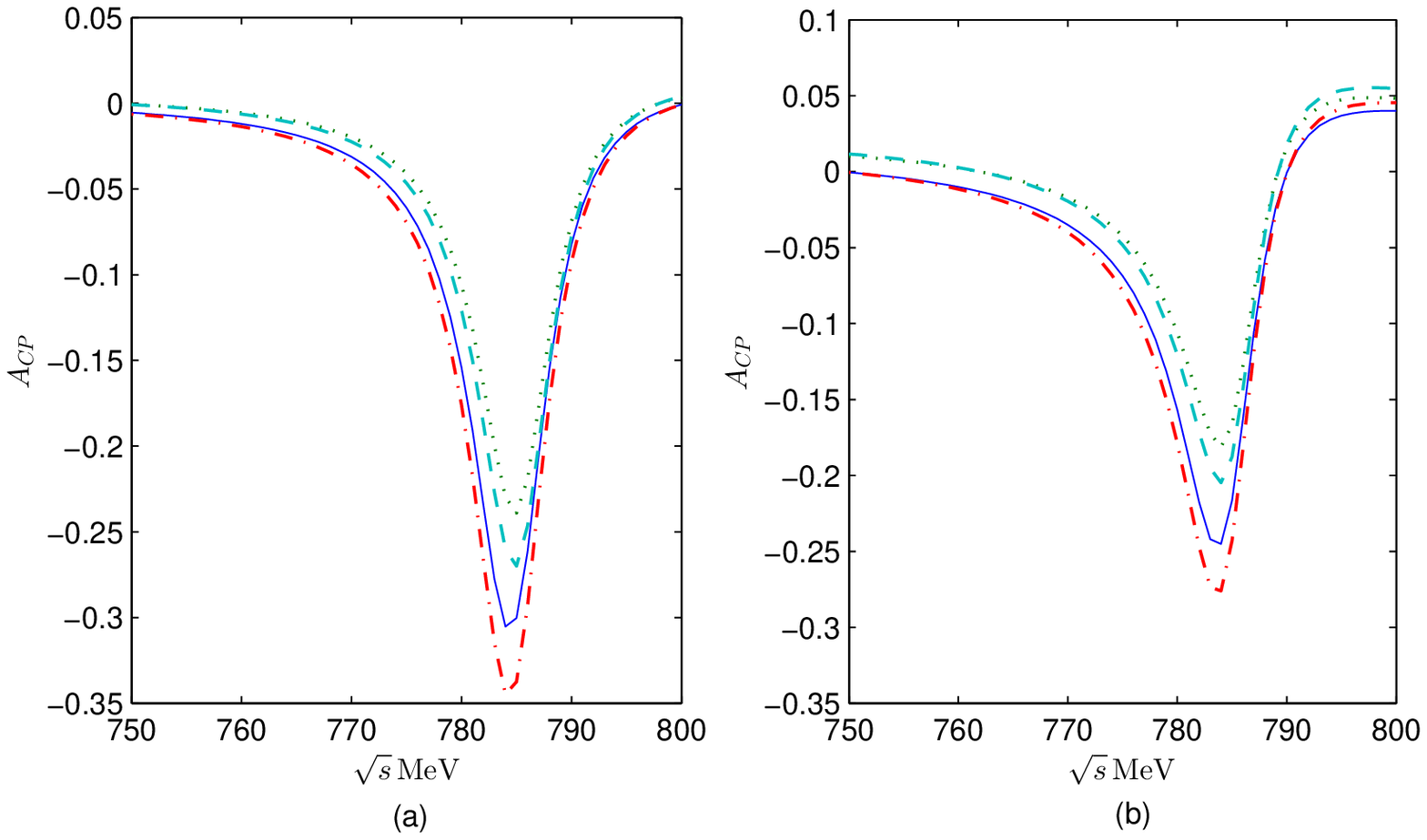}}
\centering
\caption{The differential asymmetry, $A_{CP}$, in Model 1. (a) For $q^2/m_b^2=0.3$: Solid and dot lines correspond $N_c=2.78$ and $N_c=9.68$ with maximum CKM matrix elements, respectively. Dot-dashed and dashed line correspond $N_c=2.78$ and $N_c=9.68$ with minimum CKM matrix elements, respectively; (b) Same as (a) but for $q^2/m_b^2=0.5$ and $N_c=2.85$ and 9.02.}
\end{figure*}

In our numerical calculations we have several parameters: $q^2$, $N_c$, and the CKM matrix elements in the Wolfenstein parametrization. As mentioned in Sect.~2.1, the value of $q^2$ is conventionally chosen to be in the range $0.3<q^2/m_b^2<0.5$. The CKM matrix, which should be determined from the experimental data, has the following form in terms of the Wolfenstein parameters, $A$, $\lambda$, $\rho$, $\eta$ \cite{K.A.}:
\begin{equation}
\left(
\begin{array}{ccc}
  1-\tfrac{1}{2}\lambda^2   & \lambda                  &A\lambda^3(\rho-\mathrm{i}\eta) \\
  -\lambda                 & 1-\tfrac{1}{2}\lambda^2   &A\lambda^2 \\
  A\lambda^3(1-\rho-\mathrm{i}\eta) & -A\lambda^2              &1\\
\end{array}
\right),
\end{equation}
where $O(\lambda^4)$ corrections are neglected. Since $\lambda$ and $A$ are well determined and the uncertainties due to the CKM matrix elements are mostly from $\rho$ and $\eta$, we take the central values of $\lambda(=0.225)$ and $A(=0.814)$ in the following.
The ranges for $\rho$ and $\eta$ are
\begin{equation}
\bar \rho=0.117\pm0.021,\qquad \bar \eta=0.353\pm0.013,
\end{equation}
with
\begin{equation}
\bar \rho=\rho\Bigg(1-\frac{\lambda^2}{2}\Bigg),\qquad \bar \eta=\eta\Bigg(1-\frac{\lambda^2}{2}\Bigg).\nonumber
\end{equation}
The form factors $F_1(m^2_\rho)$ and $A_0(m^2_\pi)$ depend on the inner structure of the hadrons. Under the nearest pole dominance assumption, we take the following $k^2$ dependence of the form factors:
for Model 1(2) \cite{M.B.,Y.H.2,D.C.,P.B.2,A.A.,H.Y.4}:
\begin{equation}
F_1(k^2)=\frac{h_1}{1-\frac{k^2}{m_1^2}},\qquad A_0(k^2)=\frac{h_{A_0}}{1-\frac{k^2}{m_{A_0}^2}},
\end{equation}
where $h_1=0.25(0.292)$, $h_{A_0}=0.30(0.366)$, \\$m_1=5.32\,\mathrm{GeV}$, and $m_{A_0}=5.27\,\mathrm{GeV}$;
for Model 3(4) \cite{Y.H.2,D.C.,P.B.2,A.A.,H.Y.4,Y.K.2}:
\begin{equation}
F_1(k^2)=\frac{h_1}{\big(1-\frac{k^2}{m_1^2}\big)^2},\qquad A_0(k^2)=\frac{h_{A_0}}{\big(1-\frac{k^2}{m_{A_0}^2}\big)^2},
\end{equation}
where $h_1=0.25(0.292)$, $h_{A_0}=0.30(0.366)$, $m_1=5.32\,\mathrm{GeV}$, and $m_{A_0}=5.27\,\mathrm{GeV}$;
for Model 5 \cite{P.B.1,H.B.F.}:
\begin{eqnarray}
F_1(k^2)&=&\frac{h_1}{1-x_1\frac{k^2}{m_1^2}+y_1\big(\frac{k^2}{m_1^2}\big)^2},\nonumber\\ A_0(k^2)&=&\frac{h_{A_0}}{1-x_0\frac{k^2}{m_{A_0}^2}+y_0\big(\frac{k^2}{m_{A_0}^2}\big)^2},
\end{eqnarray}
where $h_1=0.261$, $h_{A_0}=0.302$, $x_1=2.03$, $y_1=1.29$, $x_0=-1.49$, $y_0=6.61$, $m_1=5.32\,\mathrm{GeV}$, and $m_{A_0}=5.27\,\mathrm{GeV}$. The decay constants used in our calculations are $f_\rho=216\,\mathrm{MeV}$ and $f_\pi=132\,\mathrm{MeV}$ \cite{H.Y.4}. 

In the numerical calculations, it is found that for a fixed $N_c$, there is a maximum value for the differential CP violating parameter, when the low invariant mass of the $\pi^+\pi^-$ pair is in the vicinity of the $\omega$ resonance, $0.780-0.785\,\mathrm{GeV}$. Five models with different form factors were investigated to study the model dependence of $A_{CP}$ in Ref. \cite{X.H.2}. To be more specific, in Figs.~1a, b we display the results for the form factors in Model 1. These results show explicitly the dependence of the CP violating asymmetry on $q^2/m_b^2$, CKM matrix elements and the effective parameter $N_c$. The dependence on $N_c$ comes form the fact that $N_c$ is related to the hadronization effects, and consequently, we cannot exactly determine $N_c$ in our calculations. Therefore, we treat $N_c$ as a free effective parameter and take it in the range $2.07(2.09)<N_c<4.54(4.65)$ when $q^2/m_b^2=0.3(0.5)$ for reasons which will be explained later (Sect. 3).

Then we calculate the localized integrated CP asymmetries. According to Eq.~(18), we integrate $A_{CP}$ over the low invariant mass of $\pi^+\pi^-$ ($\sqrt s$) and obtain the localized integrated asymmetries $A_{CP}^\Omega$. Considering the significant region of $\rho$-$\omega$ mixing, we choose the integration interval of $\sqrt s$ to be from 0.750 to 0.800\,GeV. In order to compare with the newest result of the LHCb experiments, we also calculate $A_{CP}^\Omega$ when $\sqrt s$ is in the low-mass region ($0.470<\sqrt s<0.770$\,GeV), the high-mass region ($0.770<\sqrt s<0.920$\,GeV) and the whole region ($0.470<\sqrt s<0.920$\,GeV) near the $\rho^0$ resonance \cite{R.A.2}. The numerical results are displayed in Table 1. We also display $A^\Omega_{CP}$ with and without $\rho$-$\omega$ mixing when $0.750<\sqrt s<0.800$\,GeV and $0.470<\sqrt s<0.920$\,GeV in Table 1.

Table 1 shows that the values of $A^{\Omega}_{CP}$ in our calculations vary from -0.0752 (-0.0626) to -0.0403 (-0.0290) corresponding to $q^2/m_b^2=0.3(0.5)$, in the regions of $N_c$, the CKM matrix elements, and the form factors in five models when $0.750<\sqrt s<0.800$\,GeV. From Fig.~4 of Ref.~\cite{R.A.2}, we can see the localized integrated CP asymmetries have different signs in $\cos\theta>0$ and $\cos\theta<0$ regions when $0.750<m(\pi^+\pi^-)_{low}<0.800\, \text{GeV}$. If one adds the events in these two experimental regions together, the total localized integrated CP asymmetry obtained from experiment becomes $-0.0294\pm0.0285$ when $0.750<m(\pi^+\pi^-)_{low}<0.800\, \text{GeV}$. The values in our calculations agree with this experimental data. We stress that $A_{CP}^\Omega$ in our calculations is always negative in this integration region and its sign is independent of form factor models. We note that the signs of $A^\Omega_{CP}$ are positive when $\rho$-$\omega$ mixing is not considered in this region. This indicates that $\rho$-$\omega$ mixing is vital for $A_{CP}^\Omega$ to be negative in this region.

From Table 1, we can see $\rho$-$\omega$ mixing changes the sign of $A^\Omega_{CP}$ from positive to negative. Figs.~1(a) and 1(b) show that the peak of the differential asymmetry $A_{CP}$ involving $\rho$-$\omega$ mixing is on the right of $0.770$\,GeV. Therefore, comparing with $A_{CP}^\Omega$ in the range $0.470<\sqrt s<0.770$\,GeV, the localized integrated CP asymmetries move towards the negative direction when $0.770<\sqrt s<0.920$\,GeV due to $\rho$-$\omega$ mixing. This behavior is also independent of form factor models. In fact, in our calculations the difference between the localized integrated CP asymmetries in the regions $0.470<\sqrt s<0.770$\,GeV and $0.770<\sqrt s<0.920$\,GeV varies from 0.0076 to 0.0387. We add the experimental events for positive and negative $\cos \theta$ in the regions $0.470<m(\pi^+\pi^-)_{low}<0.770$\,GeV and $0.770<\sqrt s<0.920$\,GeV and obtain $A^\Omega_{CP}$ as $0.0508\pm0.0171$ and $-0.0256\pm0.0202$, respectively, from the data in TABLE IV of Ref.~\cite{R.A.2}. We can see they have opposite signs. After we combine these two regions together, $A^\Omega_{CP}$ in the whole region $0.470<m(\pi^+\pi^-)_{low}<0.920$\,GeV is $0.0173\pm0.0130$ \cite{R.A.2}. The values of $A^\Omega_{CP}$ with $\rho$-$\omega$ mixing shown in Table~1 differ a lot between Model 1 (or 2,3,4) and Model 5. $A^\Omega_{CP}$ is also sensitive to the choice of $q^2/m_b^2$. When $0.470<\sqrt s<0.770$\,GeV, it appears that $A^\Omega_{CP}$ varies from 0.0278 to 0.0353 in Models 1,2,3 and 4 for $q^2/m_b^2=0.5$ and $N_c=4.65$. This result is consistent with the experimental data. When $0.770<\sqrt s<0.920$\,GeV, except for $q^2/m_b^2=0.5$ and $N_c=4.65$ in Models 1,2,3 and 4, $A^\Omega_{CP}$ varies from  -0.0408 to -0.0110. This result is consistent with the experimental data. In the whole region of $0.470<\sqrt s<0.920$\,GeV, $A^\Omega_{CP}$ varies from 0.0096 to 0.0146 in Models 1,2,3 and 4 for $q^2/m_b^2=0.5$ and $N_c=4.65$. This result is also consistent with the experimental data.

From above discussions, we can see $A^{\Omega}_{CP}$ with $\rho$-$\omega$ mixing, especially the signs, agree with the experimental result when $0.750<\sqrt s<0.800$\,GeV. We also find that the localized integrated CP asymmetries move toward the negative direction due to $\rho$-$\omega$ mixing. $A^{\Omega}_{CP}$ in the region $0.770<\sqrt s<0.920$\,GeV contains the contribution of $\rho$-$\omega$ mixing while that in the region $0.470<\sqrt s<0.770$\,GeV does not. Therefore, $\rho$-$\omega$ mixing contributes to the sign change of CP asymmetry around the $\rho^0(770)$ peak of $m(\pi^+\pi^-)_{low}$. Our results in these two regions and the whole region are consistent with the experimental data for several choices of $q^2/m^2$ and models of form factors. One should take the effect of $\rho$-$\omega$ mixing into account in order to answer the question why the sign of CP asymmetry changes around the $\rho^0(770)$ peak of $m(\pi^+\pi^-)_{low}$.

\vspace{4cm}
\section{Extraction of $N_c$ from data of branching ratios}

\subsection{Formalism}
\begin{figure*}[hbt]
\centering
\scalebox{0.8}[0.8]{\includegraphics[54,268][540,564]{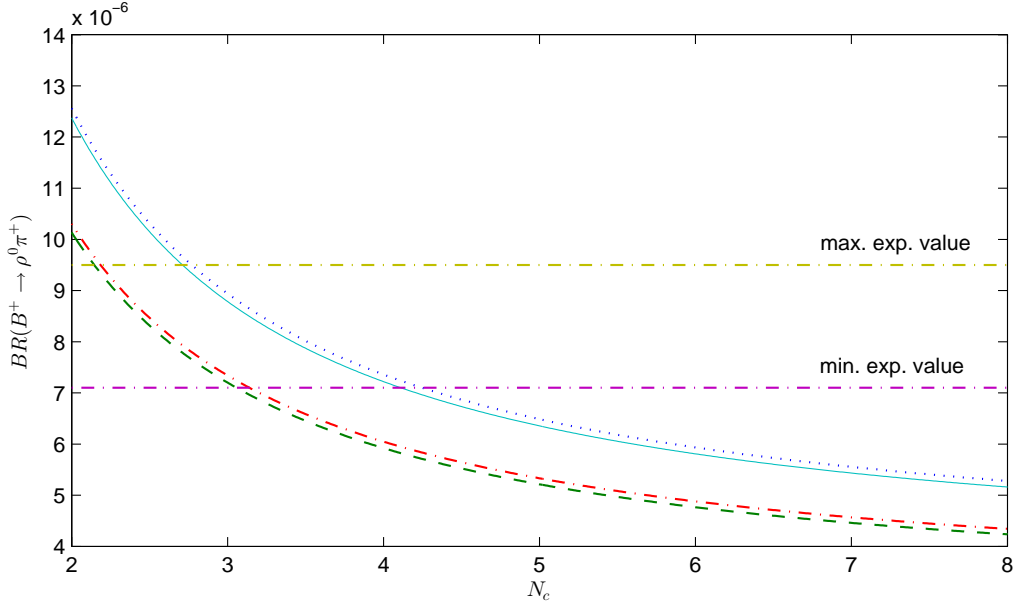}}
\caption{The branching ratio for $B^+\to\rho^0 \pi^+$ in Model 2. Solid (dashed) line stands for $q^2/m_b^2=0.3$ and maximum (miximum) CKM matrix elements. Dot (dot-dashed) line stands for $k^2/m_b^2=0.5$ and maximum (miximum) CKM matrix elements.}
\end{figure*}

As mentioned before, $N_c$ includes nonfactorizable effects which are difficult to deal with at present. Therefore, we treat $N_c$ as an effective parameter to be determined by the experimental data. With the factorized decay amplitudes, we can calculate the decay rates by using the following expression \cite{Y.H.}:
\begin{eqnarray}
\Gamma(B \to VP)=\frac{|\vec{p}_\rho|^3}{8\pi m^2_V}\Big|\frac{A(B \to VP)}{\epsilon^*\cdot p_B}\Big|^2,
\end{eqnarray}
where
\begin{eqnarray}
|\vec{p}_\rho|=\frac{\sqrt{[m_B^2-(m_V+m_P)^2][m_B^2-(m_V-m_P)^2]}}{2m_B}
\end{eqnarray}
is the c.m. momentum of the decay particles, and $A(B\to VP)$ is the decay amplitude. In our case, to be consistent, we should also take into account the $\rho$-$\omega$ mixing contribution when we calculate the branching ratio since we are working to the first order of isospin violation. Explicitly, for $B^+\to \rho^0\pi^+$, we obtain
\begin{eqnarray}
&&BR(B^+\to\rho^0\pi^+)\nonumber\\
&=&\frac{G_F^2|\vec p_\rho|^3}{16\pi m_{\rho}^2 \Gamma_{B^+}}\Big|[V_{ud}V^*_{ub}A^T_{\rho^0}-V_{td}V^*_{tb}A^P_{\rho^0}]/(\epsilon^* \cdot p_B)\nonumber\\
&&\qquad +[V_{ud}V^*_{ub}A^T_{\omega}-V_{td}V^*_{tb}A^P_{\omega}]/(\epsilon^*\cdotp_B)\nonumber\\
&&\qquad\qquad\quad\times\frac{\tilde{\Pi}_{\rho\omega}}{(s_\rho-m_\omega^2)+\text{i}m_\omega\Gamma_\omega}\Big|^2,
\end{eqnarray}
where the tree and penguin amplitudes are
\begin{eqnarray}
A^T_{\rho^0}&=&a_1X^{(B^+\rho^0,\pi^+)}+a_2X^{(B^+\pi^+,\rho^0)},\nonumber\\
A^P_{\rho^0}&=&\Big(-a_4+\frac{3}{2}a_7+\frac{3}{2}a_9+\frac{1}{2}a_{10}\Big)X^{(B^+\pi^+,\rho^0)}\nonumber\\
&&+\Big[a_4+a_{10}-2(a_6+a_8)\frac{m^2_{\pi}}{(m_u+m_d)(m_b+m_u)}\Big]\nonumber\\
&&\qquad\qquad \times X^{(B^+\rho^0,\pi^+)},\nonumber\\
A^T_{\omega}&=&a_1X^{(B^+\omega,\pi^+)}+a_2X^{(B^+\pi^+,\omega)},\nonumber\\
A^P_{\omega}&=&\Big[2a_3+a_4+2a_5+\frac{1}{2}(a_7+a_9-a_{10})\Big]X^{(B^+\pi^+,\omega)}\nonumber\\
&&+\Big[a_4+a_{10}-2(a_6+a_8)\frac{m^2_{\pi}}{(m_u+m_d)(m_b+m_u)}\Big]\nonumber\\
&&\qquad\qquad \times X^{(B^+\omega,\pi^+)}.
\end{eqnarray}

For $B^0\to\rho^+\pi^-$, we obtain
\begin{eqnarray}
&&BR(B^0\to\rho^+\pi^-)=\frac{G_F^2|\vec p_\rho|^3}{16\pi m_{\rho}^2 \Gamma_{B^0}}\nonumber\\
&&\qquad\times|(V_{ub}V^*_{ud}A^T_{\rho^+}-V_{tb}V^*_{td}A^P_{\rho^+})/(\epsilon^* \cdot p_B)|^2,
\end{eqnarray}
where
\begin{eqnarray}
A^T_{\rho^+}&=&a_1X^{(B^0\pi^+,\rho^-)},\nonumber\\
A^P_{\rho^+}&=&(a_4+a_{10})X^{(B^0\rho^-,\pi^+)},
\end{eqnarray}
where $X^{(B^0\rho^+,\pi^-)}=2f_\pi m_\rho A_0(m_\pi^2)(\epsilon^* \cdot p_B)$ and\\ $X^{(B^0\pi^-,\rho^+)}=\sqrt 2 f_\rho m_\rho F_1(m_\rho^2)(\epsilon^* \cdot p_B)$.

\subsection{Numerical results}
\begin{table*}[!hbt]
\centering
\caption{Summary of the ranges of $N_c$ which are determined from the experimental data for various models and input parameters [numbers outside (inside) brackets are for $q^2/m_b^2=0.3(0.5)$]. The notation (number, number) means the lower and upper limits for $N_c$. (-, -) means that there is no range of $N_c$ which is consistent with the experimental data.}
\begin{tabular*}{14cm}{@{\extracolsep{\fill}}cccc}
\hline
\hline
&$B^+$& $B^0$\\
\hline
Model~1\\
$\rho_{max},\eta_{max}$    &(2.70, 4.09) [(2.76, 4.22)]  &  (-, -) [(-, -)]     \\
$\rho_{min},\eta_{min}$    &(2.14, 3.05) [(2.18, 3.13)]  &  (-, -) [(-, -)]      \\
Model~2\\
$\rho_{max},\eta_{max}$     &(-,-) [(-,-)]  &   (-,-) [(-,-)]     \\
$\rho_{min},\eta_{min}$     &(-,-) [(-,-)]  &   (-,-) [(-,-)]     \\
Model~3\\
$\rho_{max},\eta_{max}$    &(2.24, 3.23) [(2.28, 3.32)]  &  (0.05, 0.07) [(0.10, 0.14)]     \\
$\rho_{min},\eta_{min}$    &(2.85, 4.40) [(2.92, 4.55)]  &  (0.06, 0.13) [(0.12, 0.15)]      \\
Model~4\\
$\rho_{max},\eta_{max}$     &(-,-) [(-,-)]  &   (-,-) [(-,-)]     \\
$\rho_{min},\eta_{min}$     &(-,-) [(-,-)]  &   (-,-) [(-,-)]      \\
Model~5\\
$\rho_{max},\eta_{max}$    &(2.71,4.54) [(2.76,4.65)]  &   (0.11,0.12) [(0.11,0.12)]     \\
$\rho_{min},\eta_{min}$    &(2.07,3.13) [(2.09,3.18)]  &   (0.10,0.11) [(0.10,0.11)]      \\
\hline
\hline
\end{tabular*}
\end{table*}

The latest experimental data of branching ratios of $B^+\to\rho^0\pi^+$ and $B^0\to\rho^+\pi^-$ from Particle Data Group (PDG) are \cite{K.A.}
\begin{eqnarray}
BR(B^+\to\rho^0\pi^+)&=&(8.3\pm1.2)\times10^{-6},\nonumber\\
BR(B^0\to\rho^+\pi^-)&=&(2.3\pm0.23)\times10^{-5}.\nonumber
\end{eqnarray}

We can determine the range of $N_c$ by comparing the theoretical values of the branching ratios with those of two-body decay channels. We calculate the branching ratios with the formula given in Eqs.~(32), (34) and (36) in five models for the weak form factors which are mentioned in the previous subsection. In Fig.~2, we show the results for $B^+\to\rho^0\pi^+$ in Model 1 as an example. The numerical results are sensitive to uncertainties coming from the experimental data. In addition, the branching ratio also depends on the CKM matrix elements which are parameterized by $\lambda$, $A$, $\rho$, and $\eta$. In the allowed ranges for the parameters $\rho$ and $\eta$, we obtain the range of $N_c$. We summarize the allowed range of $N_c$ in Table 2. It is found that the experimental data constrain the value of $N_c$ into two regions. We note that if $N_c$ approaches to zero, nonperturbative effects would be very large. Considering this, we drop the range $N_c<1$. Therefore, $N_c$ could be in the range $2.07(2.09)<N_c<4.54(4.65)$ for $q^2/m_b^2=0.3(0.5)$. These values have been used in Sect. 2.

\section{Conclusion and discussion}

The first aim of the present work is to study the localized integrated CP asymmetry for the decays $B^\pm\to\rho^0(\omega)\pi^\pm\to \pi^+\pi^-\pi^\pm$ with the inclusion of $\rho$-$\omega$ mixing. The second aim is to study the sign change caused by $\rho$-$\omega$ mixing.

In the calculation of CP violating asymmetry parameters, we need the Wilson coefficients for the tree and penguin operators at the scale $m_b$. We worked with the renormalization scheme independent Wilson coefficients. One of the major uncertainties in our calculations is due to the fact that hadronic matrix elements of both tree and penguin operators involve nonperturbative QCD effects. We worked in the factorization approximation, with $N_c$ being treated as an effective parameter to include nonfactorizable contributions. We compared our theoretical results with the latest experimental data from PDG to determine the range of $N_c$ as $2.07(2.09)<N_c<4.54(4.65)$ for $q^2/m_b^2=0.3(0.5)$. It has been pointed out that the factorization approach is quite reliable in energetic weak decays \cite{H.Y.1}.

We explicitly showed that the CP violating asymmetry is very sensitive to $N_c$, the CKM matrix elements and the form factors. There is a maximum value for the differential CP violating parameter when the low invariant mass of the $\pi^+\pi^-$ pair is near the vicinity of the $\omega$ resonance, $0.780-785\,\mathrm{GeV}$. We determined the range of the localized integrated CP asymmetry with and without $\rho$-$\omega$ mixing in the ranges of $N_c$, the CKM matrix elements, and $q^2/m_b^2$. For all the models investigated, we found that the localized integrated CP violating asymmetry with $\rho$-$\omega$ mixing varies from -0.0403 (-0.0290) to -0.0752 (-0.0626) corresponding to $q^2/m_b^2=0.3(0.5)$ when $0.750<m(\pi^+\pi^-)_{low}<0.800\, \text{GeV}$. If one adds the events in the $\cos \theta<0$ and $\cos \theta>0$ experimental regions together when $0.750<m(\pi^+\pi^-)_{low}<0.800\, \text{GeV}$, the total localized integrated CP asymmetry is $-0.0294\pm0.0285$. Our results, especially the signs, agree with the experimental data when $0.750<m(\pi^+\pi^-)_{low}<0.800\, \text{GeV}$. We note that the signs are positive in this region when $\rho$-$\omega$ mixing is not considered. This indicates that $\rho$-$\omega$ mixing is vital for $A_{CP}^\Omega$ to be negative in this region. It was shown that the maximum localized integrated asymmetry in the range $0.750<\sqrt s<0.800$\,GeV can reach -0.0752. We also found that $\rho$-$\omega$ mixing can make the localized integrated CP asymmetries move toward the negative direction, and therefore contributes to the sign change around the $\rho^0(770)$ peak of $m(\pi^+\pi^-)_{low}$. This behavior is independent of the form factor models. In our calculations, the difference between the localized integrated CP asymmetries in the regions $0.470<\sqrt s<0.770$\,GeV and $0.770<\sqrt s<0.920$\,GeV varies from 0.0076 to 0.0387. Our results by including the $\rho$-$\omega$ mixing mechanism in these two regions and the whole region around the $\rho^0(770)$ peak are consistent with the experimental results for some models of the form factors.

At this stage, we cannot explain the LHCb experimental data in the regions of positive and negative $\cos \theta$ individually. This is because three-body decays of heavy mesons are more complicated than two-body decays as they receive more contributions from different mechanisms, for example, nonresonants \cite{H.Y.}, the interference between intermediate resonances and final-state $KK\longleftrightarrow\pi\pi$ rescattering. We will investigate the angle distribution of $A_{CP}^\Omega$ when considering both the $\rho$-$\omega$ mixing mechanism and the interference between different spin intermediate resonances \cite{Z.H.}. We will also apply more accurate data in the future to further decrease the uncertainties in the calculations. With parameters with smaller uncertainties, we expect to be able to obtain the effects of $\rho$-$\omega$ mixing more precisely. This is important to interpret the angle distribution and the sign change of the CP asymmetry around the $\rho^0(770)$ peak of $m^2(\pi^+\pi^-)_{low}$ more accurately.

\section{Acknowledge}

This work was supported in part by National Natural Science Foundation of China (Project Nos. 11175020, 11275025, 11447021, and 11575023).

\end{document}